# How one can make the bifurcation of Maxwell's demon

# in Granular Gas   Hyper-Critical

## P. Evesque

**Lab MSSMat,  UMR 8579 CNRS, Ecole Centrale Paris**
**92295 Châtenay-Malabry, France,** e-mail: **pierre.evesque@ecp.fr**

### Abstract:

*Experimental data from Maxwell's demon experiments on granular gas are revisited. It is shown that the transition nature changes from critical to sub-critical via a tri-critical point when frequency of vibration is increased continuously. So (i) the transition is not hyper-critical as asserted previously, but (ii)  the fluctuations amplitude is spontaneously reinforced at the tri-critical point compared to the one at other frequencies. So, (iii) this analysis still contradicts a recent study which asserts that the bifurcation is always critical, and that  fluctuations shall depend on the number of grains only. Beside, (iv) it is proposed a way to build an experiment, based on some modification of the "Maxwell's demon in granular gas" set up, which undergoes a hyper-critical bifurcation, with the use of some controlled feed-back of the flows. This last idea can be generalised and used in other domains where bifurcation theory applies, i.e. physics, chemistry, population dynamics, economy,..., to improve regulation , but generating  hyper- fluctuations. Finally it is shown that the dynamical  result is sensitive to the very-details of the regulation.*

"Maxwell's demon" in granular gas is a strange effect which occurs in 1g when vibrating vertically two closed boxes connected via a horizontal slit and filled with some amount of grains. In this case one observes the filling of one box and the emptying of the other one in some range of experimental parameters.

In the case of granular gas, the state of which will be defined a little later, these parameters are at least the slit position h and width w, the vibration amplitude A and frequency f, the dissipation characteristics during ball-ball collisions and ball-wall collisions and the total number of grains $N_1+N_2$ ($N_1$ and $N_2$ refer to the number of grains in boxes 1 and 2 respectively). It can be shown that equi-repartition, *i.e.* $N_1=N_2$, is obtained at high excitation and that the population difference $N_1-N_2$ varies with A & f  below some threshold.

Stated as this, the problem looks "strange" but "simple". However, things are more complicated actually. Indeed, this effect is a good example on how sensitive and how complex the behaviour of granular matter can be. This will be summed up in the two next paragraphs:

Consider the 2 boxes **filled with less than two layers** of grains; in this case the grains move as a disperse gas with random collisions when system is shaken; this is what we will call the "granular gas" state. Then, in this case, the Maxwell's demon effect is observed due to the increase of the number of bead-bead collisions and of the energy dissipation it generates [1]; it occurs mainly since the dissipation increases non linearly with the number density of grains in the cell. Also the width w of the slit has some importance [2]: if it is smaller than the mean free path $l_c$ between two ball-ball collisions, then the system is controlled by the flow function J(N) of balls flowing from one box to the other one; in this case J(N) depends only on the number N of balls in the box they are leaving, and not on the ones of the box to which they are flowing. Conversely, when w> $l_c$, one has to consider equilibrium of pressure p between the two boxes at the level of the slit, because collisions can occur in the vicinity of the slit; this modifies the rules of flow equilibrium.

However other experiments exist where one can also observe some generation of overpopulation in a box compared to the other one. This occurs for instance with two cells containing **few centimetres /layers of grains each [3-5].** Where one sees the emptying of a box and the filling of the other one above some threshold of vibration. So the behaviour looks quite similar to the "Maxwell's demon in granular gas", but in a different range of parameters and with different laws. In this case, the mechanism(s) is (are) different: among others, a possible mechanism is the effect of the ambient air which can provoke an irreversible pumping [6-8] which acts intermittently when the two piles are pulled up (*i.e.* lifted up from the bottom plate in the box frame); this occurs periodically at some phase of the vibration. This occurs because the granular matter acts as a porous through which air has to flow. The motion generates a pumping which is proportional to the difference of grain height in each box and it is caused by the permeation of the air through the porous medium, and the motion of the grains. At an other phase of the period, the grains land on the bottom and the system becomes solid-like, blocked by solid friction, so that the flow of grains occur always in the same direction intermittently. Of course, the pumping difference is negligible when the two levels are equal, but it becomes larger and larger after the mechanism has started, *i.e.* as soon as the difference of level is increasing, *i.e.* as soon as the difference starts being generated, so that one observes a spontaneous symmetry breaking. This effect disappears when using vacuum instead of air and when the height of grains is low enough so that depression linked to permeation becomes negligible. Other effects can probably be invoked such as the "pseudo-shovel effect" [3] , which requires also a biased motion of a lateral wall. Few examples of such behaviours can be found in video [4] or in the literature [5].

Coming back to the "Maxwell's demon effect in granular gas" more specifically, the problem is even more complex than it is told was in previous paragraphs. To simplify, we restrict the discussion in the case when the slit width is small compared to the mean free path $l_c$ so that the physics is controlled by the flows $J_1(N_1)$ and $J_2(N_2)$ from each box, which was the case assumed in [1]; so we omit discussing the effect a possible feed back linked to collisions occurring betweens balls flowing out of cells 1 & 2 in the vicinity of the slit and to functions of two variables $J_1(N_1,N_2)$.

In fact, as shown in [9], $J_{1or2}(N)$ can be determined easily from an experiment with a single box, via the study of its emptying ; this allows the direct measure of J(N). And it is easily found [9] that the model proposed in [1] does not fit experimental data on J(N). In particular, as shown in [9] the flow from the outlet increases linearly with the number N of balls near N=0 (instead of $N^2$ as predicted in [1]). This linear dependence may perhaps be due to some lack of sphericity of the balls or to some horizontal vibration. Anyway, this is an experimental evidence which is now recognised by others [10] ; it proves that the prediction from theoretical approach have to be verified.

Besides that in [1] and in the other papers related to the topics, the bifurcation which is undergone in "Maxwell's demon in granular gas" has been always supposed to be critical. This is linked to the peculiar function J(N) used in [1] which forces the bifurcation to be critical. As we have just shown that this function is not the correct one [9], this requires to study the complete bifurcation for different possible J(N). This has been done in [11], and it has been found that, depending on the shape J, the bifurcation can be either critical or sub-critical if the wing before the maximum (left wing) is steeper than the one after the maximum (right wing) or the inverse. Furthermore a new kind of bifurcation, which has been denominated "hyper-critical", has been found in the case when J is symmetric. These points are recalled in the appendix where the dynamics of the two-box system is also determined from the function J(N). Due to this, the true shape of the flow function J(N) was determined in [11], and it was found that it evolves with the parameters (A,f) of vibration so that the bifurcation passes from critical to sub-critical when increasing the frequency of vibration. We have concluded a little too fast that it could become hyper-critical. This is this point that we want to discuss in this paper. We will show (i) that there is no reason to claim that the bifurcation is "spontaneously" hyper-critical when adjusting the frequency to the one which makes the transition between a critical to a sub-critical bifurcation. However, we will discuss also (ii) whether it is possible or not to make the bifurcation hyper-critical by controlling the flow using some feed-back. We will show that this is possible.

## Why the bifurcation encountered in the experiment of "Maxwell's demon in granular gas" is not hypercritical.

We report in Fig. 1 typical flow curves J(N) for different frequency of vibration (data are taken from [11]). The study of the bifurcation is recalled in the Appendix. In short, (i) the bifurcation is shown to occur at N=$N_{max}$, *i.e.* when J is maximum, and/or dJ/dx=0, (ii) the bifurcation is found to be hyper-critical when the flow function J(N) is symmetric compared to $N_{max}$, *i.e.* when J($N_{max}$-k)= J($N_{max}$+k) whatever k. This symmetry condition imposes $d^{2n+1}J/dx^{2n+1} = 0$ at $N_{max}$, whatever n>0. We note that result (i) says already that dJ/dx=0 at $N_{max}$, which is the condition to be at the threshold of the bifurcation.

On the other hand, the **experiment on Maxwell's demon** shows (see Fig. 1) that the flow J(N) is such that its left wing is steeper than the right wing below f=36 Hz and less steep than the right wing above f = 36 Hz. This fact (non symmetry of J) imposes simply that the first order n>1 , say $n_o$, for which $d^{2n_o+1}J/dx^{2n_o+1} \neq 0$ at f $\neq$ 36 Hz , is such that it is negative, *i.e.* $d^{2n_o+1}J/dx^{2n_o+1} < 0$, at f > 36 Hz and positive, *i.e.*

$d^{2n_o+1}J/dx^{2n_o+1} > 0$, at $f < 36Hz$, so that it changes of sign at $f=36Hz$. The order $n_o$ which is concerned in the present experiment is likely $n_o=1$. We have not enough accuracy to measure the higher order terms $(d^{2n+1}J/dx^{2n+1})_{N_{max}}$, for $n>n_o$. The experimental result does not imply of course that all odd derivative of J, $(d^{2n+1}J/dx^{2n+1})_{N_{max}}$, are 0 at $f = 36Hz$. *So, it does not imply that the bifurcation is hypercritical* (all $(d^{2n+1}J/dx^{2n+1})_{N_{max}} = 0$ at $f = 36Hz$).

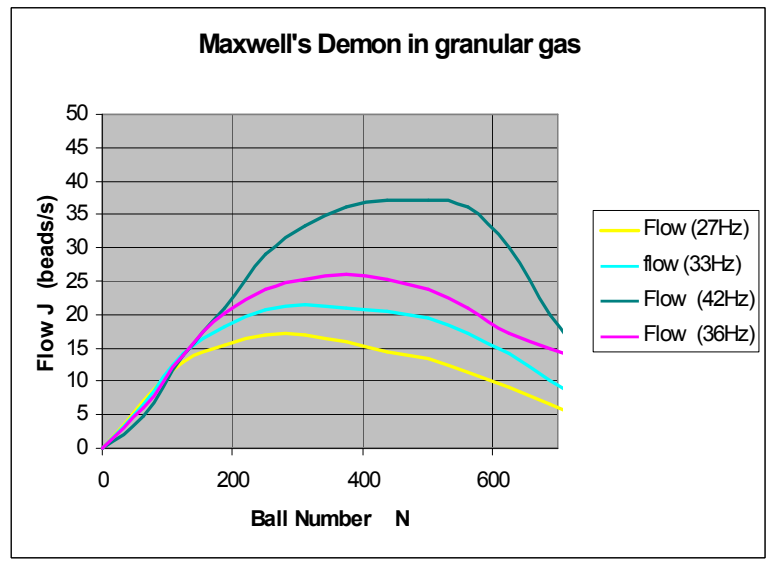

**Figure 1:** *When a box contains N balls (diameter d) and that a horizontal slit is made at some height h with some width w, the grains flow out from the box when the box is vibrated vertically (amplitude A, frequency f). Typical flow curves as a function of N at different frequencies (27-33-36-42Hz) from [11]; the other parameters are: A=1.25mm,d=1mm, h=10-13mm, w= 3mm, and the box size is: H= 30mm, S=20\*12mm². This flow curves can be used to study the distribution of bead population in two boxes connected via a slit and vibrated vertically in the limit of non interacting flows(see appendix).*

On the other hand, this kind of behaviour at $f=36Hz$ is well known in bifurcation theory and/or in phase transition: it is the analogue of a tri-critical point [13], for which one knows that fluctuations are enhanced. Indeed convergence towards the equilibrium dynamics is only forced by terms $(d^{2n+1}J/dx^{2n+1})_{N_{max}}$ in the development of J which are of higher order nearby 36Hz than beside (*cf.* Appendix). So, one should observe less efficient convergence and larger fluctuations at this working point (36Hz). However this fluctuation enhancement is not sufficient to get the true diffusion problem among all the values of $N_1$, as predicted by the hypercritical bifurcation, this would require that all $(d^{2n+1}J/dx^{2n+1})_{N_{max}} = 0$ exactly; (and this has not been proven experimentally by the experiment as told already). The domain of N which is reachable due to fluctuations is simply larger at 36 Hz, but it is still limited to a given zone limited to the vicinity of $N_{max}$; it does not extend to N=0.

This discussion is worth to be done, because the main goal of dynamical system theory is to analyse the physics of systems in a subset of a continuous space of parameters and to learn about the physics from these observations when varying these parameters. So it is essential to assume continuity. In the present case, the right approach is then to assume that J(N) depends continuously of the parameters {A, f,…} and to learn from the experiment how J varies assuming continuous mapping, or projection of continuous mapping. It was then not correct to conclude to the hyper-criticality, which is a very specific situation, much too specific compared to what we are able to prove. Doing so we were merely denying continuity.

Nevertheless, part of the conclusion which concerns the enhancement of fluctuations at this working point is correct, so that one should be able to measure larger fluctuations at 36 Hz than nearby in this experiment. The manner to calculate/predict these fluctuations is explained in [11]; it will not be reproduced here.

It is worth mentioning that we can proceed in the same ways with the data reported in [10]. Obviously, the shape of the flow curves of Fig. 5 in [10] exhibit trends similar to the ones in Fig.1, but the frequency of transition is 55 Hz instead of 36 Hz in our case. The right wing is still steeper (less steep) than the left one at f>55 Hz (f<55 Hz). So the point at 55 Hz in [10] is also a tri-critical point; fluctuations there shall be larger than beside contrarily to what has been asserted in [10]. Such a result is also well-known in physics of critical phenomena.

It has been often reported/argued that "Maxwell's demon in granular gas" is a simple example where to apply dynamical system theory and theory of phase transition; so, it can be viewed as an archetype of the efficiency of applying these theories. Furthermore, this example, because it is simple, can be modified, complicated, coupled with other effects, so that it can lead to a series of extension [12] for which the physics has to be understood. This is why its analysis has to be as correct as possible, and why we have thought this remark should be done. So the true working point which is characterised by the first $(dJ/dN)_{N_{max}}$ and third $(d^3J/dN^3)_{N_{max}}$ derivatives is equal to 0 (while it is not 0 at a different frequency) is not the sit of a hyper-critical bifurcation, **but only of a tri-critical point**, **where fluctuations are still enhanced** compared to beside it but not as large as predicted by hyper-criticality.

An other question remains then: can one observe a hyper-critical bifurcation in some case; or is it very unlikely. For instance, can it be obtained in the case of "Maxwell's demon in granular gas" using some kind of modification? And the corollary questions are: how and when this can be observed? If we are able to make this change in the case of Maxwell's demon, this makes the result quite interesting:

Indeed, as "Maxwell's demon in granular gas" can be considered as an archetype, its behaviour can be generalised to other systems and other domains (physics, chemistry,…) and can be used to illustrate other cases. This makes these results more interesting.

## How to make hypercritical the bifurcation in "Maxwell's demon in granular gas" using an adequate control of the flow.

We still consider the limit when the two flows from each box do not interact in the vicinity of the slit for sake of simplicity. And we consider the experimental case described above, with the flow rules described by Fig. 1.

Let us now assume that the experiment can be modified in the following manner: the two compartments are now connected to two reservoirs through two tubes each; the first one is located on the top of each compartment and the other one at the bottom; the first one allows to feed with grains each compartment and the other one to empty them. It is supposed also that these flows can be enslave to the result from a computer and that we are able to measure in real time the number of grains in a compartment and the total number of grains which have flown from the tubes, so that the experiments works with a definite total number of grains $N_1+N_2$ in the two compartments. This means that the experiment allows now to generate additional flows $J'_1$ and $J'_2$ to compartment 1 and 2, which are independent from the flows $J_1$ and $J_2$ from the slit. Both $J'_1$ and $J'_2$ can be positive or negative and adjusted at will.

Dynamics can be written:

$$dN_1/dt = J(N_2) - J(N_1) + J'_1 \qquad\qquad (1.a)$$

$$dN_2/dt = J(N_1) - J(N_2) + J'_2 \qquad\qquad (1.b)$$

Constant number of grains $N_1+N_2$ is enforced imposing $J'_1+J'_2=0$ since $d(N_1+N_2)/dt = J'_1+J'_2$. As we have introduced 2 parameters, *i.e.* $J'_1$ and $J'_2$ and since grain preservation define a single constraint only, it remains one parameter to fix. This one can be chosen using Fig. 1 to enforce that $J(N_2) - J(N_1) + J'_1 = J(N_1) - J(N_2) + J'_2$ . It reads $2[J(N_2) - J(N_1)] = J'_2 - J'_1$, or since $J'_2 + J'_1 = 0$:

$$J'_2 = -J'_1 = J(N_2) - J(N_1) = J(N_{tot} - N_1) - J(N_1) \qquad\qquad (2)$$

This allows to impose also:

$$d(N_2 - N_1)/dt = 0 \qquad\qquad (3)$$

So this defines a method to *enslave the dynamics of populations to be just at equilibrium* whatever the difference $N_2-N_1$ is under a set of conditions which are (i) imposing $N_2+N_1=N_{tot}$ , (ii) measuring $N_2$ (or $N_1$) and (iii) knowing the flow curve $J$ of Fig. 1. In this case, the method allows imposing the correct value of $J'_1 = -J'_2$ at any time so that the system will work at equilibrium whatever the difference $N_2-N_1$ . It looks then at working just at criticality all the time whatever $N_2-N_1$. The systems works then as a hypercritical one. This means in particular that $N_2$ and $N_1$ will vary with time according to the fluctuations of flows. Hence it will obey some random walk in the possible domain of $N_1$ and $N_2$ . This random walk will be imposed by the randomness of the flows. In this sense the system will look hypercritical.

However, it is worth noting that the dynamics of the system and especially the dynamics of the numbers $N_1$ and $N_2$ , will strongly reflect the precise way the regulation is performed though the regulation flows $J'_1 = - J'_2$. This is because the distribution of configurations exhibited by a hyper-critical system depends strongly on its noise [11]. For instance, one shall also add some noise generator to the flows J' with a noise which obeys the noise of the flow J(N) if one wants the dynamics to respect exactly the dynamics of the hypercritical bifurcation; otherwise the statistical distribution of states will be different in the two cases. On the contrary, using a regulation J' with no noise, changes the equilibrium distribution of states; in the same manner using a flow generator injecting noise on $J'_1$ and $J'_2$ larger than those of J will lead to different distribution of equilibria.

In other words, ***the system and its behaviour become quite sensitive to parameters which are uncontrolled often***; and the dynamics of the system become controlled by little perturbations/details/noise of the regulation. This point turns out to be a major consequence as pointed out a little later.

Developing new other possible regulation which give similar trend is not difficult; it requires just to analyse the physics: Indeed, different modifications of the experiment can be proposed to obtain the "same" results, *i.e.* hypercriticality. However each of them may lead to different density of probability depending on the real nature of the control feed-back. We give two other examples here after:

► An other possible modification of the experiment could be simpler. It is obvious that the flow from a compartment varies not only with the number of grains it contains but also on the shape and size of the container, so that one can probably adjust the flow by adjusting the width $l_1$ of the box. In this case the flow from a box with $N_2$ particles can be equal to the flow from the other box with $N_1$ particles but with an other width. In this case the two boxes remain in equilibrium whatever the difference $N_1$-$N_2$ by knowing these variations and enslaving $l_1$ and $l_2$ in the right way if we know how to adjust $l_1$. It remains just measuring the number $N_1$ and adjusting in real time the with $l_1$ and $l_2$ of each cell to impose equalities of flow $j_1=j_2$.

► It is also likely that the height h between the basement of the container and the slit allows to adjust J. So, a third possibility could be to control these heights $h_1$ and $h_2$ to enforce $J_1=J_2$ whatever $N_1$ and $N_2$, $N_1+N_2$ remaining equal to $N_{tot}$.

As a conclusion, we have proved in this paper that "Maxwell's demon in granular gas" can be controlled in such a way to impose hyper-fluctuations. This is an important result because it overpasses the scope of the Maxwell's demon case, since this problem is an archetype . So the real conclusion means that any system which exhibits a bifurcation which varies within some range of parameters can be controlled easily to enforce hyper-critical behaviour. But the corollary of this is that the dynamics of the system becomes controlled by the very details of the regulation such as the noise. If not well controlled, these details may push the system to evolve erratically.

As stated above these conclusions should apply to any domain where dynamical system theory should apply, *i.e.* sociology, physics, biology, evolution of population,

econo-physics,…; so it is a very general result. As an other domain of application of dynamical system theory is economics, this paper is perhaps an explanation of what occurs in economy where people, agents, states… try to regulate, but where these actions remain sometimes/often quite surprising… and unpredictable. This work predicts that the result depends on the detailed regulation and not on the rough balance, as it is thought at first sight.

***Acknowledgements:*** CNES is thanked for partial funding. Paul Manneville is gratefully thanked for careful reading of [11] and for helpful discussion from which this paper has been built partly.

## A- Appendix: Recalls:

Indeed the problem of the "Maxwell's demon" which is observed in vibro-fluidised granular gases seems to be a good archetype for illustrating the efficiency of the theory of dynamical system. First, it is quite intriguing and spectacular, second it is quite non linear: why does one observe the filling of a half box and the emptying of the other one, when one vibrates the two half containers at an intensity smaller than a threshold acceleration $\Gamma_c$? How does the effect (and the threshold) depend on the number of grains in each compartment? On the vibration acceleration $\Gamma=a\omega^2$ or on its velocity $a\omega$? What is the dependence on the compartment size $L_1* L_2$, on the wall height h? On the bead diameter d, on its restitution coefficient $\varepsilon=v_1/v_2$?

All these questions have not yet get an answer till now; first a parallel has been tentatively drawn with a phase transition [1,2,9,10,11], probably because the observed bifurcation at $\Gamma< \Gamma_c$ looks critical and that the parallel with a second order phase transition can be addressed. But if such a parallel is valid, one shall go further in this investigation and find what are the critical exponents of the transition? And if these exponents fall into known classes?

In this appendix we develop the theoretical understanding of the Maxwell's demon in granular gas in the language of Dynamical system theory. Paper [1] has shown that the right parameter here is the flow $j_{i\rightarrow j}(N_i)$ from box i to box j. Limit of validity of this assumption is discussed in [2]. This assumption states that the flow depends on the internal characteristics of the box i from which the grains go out only, and that $j_{i\rightarrow j}$ does not seem to depend on the characteristics of box j. The validity of this assumption is related to flow passing through the slit, which has to remain small: indeed the density of beads in each compartment is small, so that the density of particles near the slit is also low and the probability of collisions between two beads is small; hence the two flows do not interact in the slit.

In [1] , the flow j has been assumed to obey some equation; but it is then easy to study the flow j(N) from a half compartment when it is disconnected from the other compartment; this needs only to determine the time dependence of its emptying. So it has been found in [9, 11] that the equation for j used in [1] is not correct. Anyway, Figure 1 gives the function j for different acceleration and as a function of N ; it shows that this flow passes through a maximum $j_{max}$ at $N=N_m$; both $j_{max}$ and $N_m$ depend on the vibration parameter. The presence of this maximum generates the bifurcation.

Indeed the evolution of the distribution obeys:

$$dN_1/dt = -j_{1\rightarrow 2}+ j_{2\rightarrow 1} \qquad \& \qquad dN_2/dt = j_{1\rightarrow 2} - j_{2\rightarrow 1} =-dN_1/dt \qquad (A1)$$

Population evolution occurs because the flows $j_{1\rightarrow 2}$ & $j_{2\rightarrow 1}$ of particles through the slit are different ; equilibrium occurs when the two flows are equal. Summing the two Eqs. (A1) leads to $d(N_1+N_2)/dt=0$, which implies: $N_1+N_2$=constant= 2N. At equilibrium $dN_1/dt=0=dN_2/dt$ which implies $j(N_1)=j(N_2)$. The equilibrium conditions

$$N_1+N_2=\text{constant}= 2N \qquad\qquad\qquad (A2.a)$$

j($N_1$)=j($N_2$) (A2.b)

**N is the mean number of particles in each box. $N_m$ is the point where j is maximum**. When $N<N_m$ it exists only a solution, *i.e.* $N_1=N_2=N$; it is stable because $-j_{1\rightarrow2}+j_{2\rightarrow1}$ is of the sign of $N_2-N_1$. But when $N>N_m$, this solution becomes unstable because the slope dj/dN is negative. We will describe in further details the physics of the transition. Obviously the flow j is fluctuating in time, even if one can introduce a mean. Nevertheless, in order to understand what is occurring we consider only that j does not fluctuate; then one will need to introduce the effect of the fluctuations to treat the problem completely. The method of treating fluctuations is recalled in [11].

## A.2 Stability analysis near the maximum of j: critical vs. sub-critical bifurcation

In the next paragraphs we show using Eq. (A1) how the bifurcation depends on the shape of j(N) near the maximum, and we relate especially the bifurcation nature to the symmetry of this curve. This is why we use a Taylor expansion of j(N) at $N=N_m$ and show how the equilibrium is modified by a change of the coefficients of this expansion. Then, we proceed in a more regular way and analyse the data as usual (in the Theory of Dynamical Systems) by expanding j(N') in the vicinity of j(N). Both are related because each set of coefficients can be expressed with the set of the other expansion.

### - Effect of the symmetry of j(N) on the kind of bifurcation

Using Taylor's series around $N_m$, one can write $j_i$(N) in the vicinity of the maximum as:

$$j_i=j_{max}+b_2(N_i-N_m)^2+b_3(N_i-N_m)^3+b_4(N_i-N_m)^4+\ldots \quad \text{(A3.a)}$$

where i stands for 1 or 2 and labels the compartment. The coefficients $b_k$ are related to the successive derivative $j^{(k)}(N_m)$ of j at $N=N_m$ according to:

$$b_k=(1/k!) \, (d^kj_i/dN^k)_{at \, N=N_m} =(1/k!) \, j^{(k)}(N_m) \quad \text{(A3.b)}$$

As j is maximum at $N_m$, $b_2$ is always negative. One can change of variable and write $x_i=(N_i-N_m)$, $u=x_1-x_2$ and $v=(x_1+x_2)$; so, v controls the total number of particles 2N since $v=2N-2N_m$; it is positive if $N_1+N_2>2N_m$ or negative if $N_1+N_2=2N<2N_m$. u defines the difference of particles between the two compartments. Inserting $j_i$ in Eq. (A1) leads to the 2 equations for the dynamics:

$$dv/dt=0 \quad \text{(A4.a)}$$

$$du/dt = -u \, [2b_2v+ b_3(3v^2+u^2)/2+b_4v(u^2+v^2)+\ldots] \quad \text{(A4.b)}$$

or

$$du/dt = -u \, [2b_2v+ 3b_3v^2/2+ b_4v^3 + u^2\{b_3/2+b_4v\}+\ldots] \quad \text{(A4.b)}$$

Solution of Eq. (A4.a) is:

$$N_1+N_2=2N_m+v \quad \text{(A5)}$$

Stationary solutions of Eq. (A4.b) are solutions of

$$0= u \, [2b_2v+ 3b_3v^2/2+ b_4v^3 + u^2\{b_3/2+b_4v\}+\ldots]$$

As $b_2<0$, the solution u=0 is stable when v<0 in the vicinity of $N_m$, *i.e.* when $v\rightarrow0^-$ since (1/u)du/dt<0. This gives $N_1=N_2=N<N_m$. Or

$$u_i=0 \quad \text{(A6.a)}$$

This solution becomes unstable when v>0. The one which is stable if $b_3>0$ is twofold:

$$u_\pm=\pm(-4b_2v+3b_3v^2+2b_4v^3)^{½}/(b_3+2b_4v)^{½}$$

which becomes when $v\rightarrow0^+$ to leading order in v:

$$u_\pm \approx \pm(-4b_2v/b_3)^{½} \quad \text{(A6.b)}$$

So $u_{\pm} \to 0$ when $v \to 0^+$ as $v^{1/2}$. So u varies continuously in the vicinity of $N_m^+$; but the slope of u at $N_m$ changes abruptly from horizontal, since u=0 for v<0, to vertical, since $u_{\pm} \approx \pm(-4b_2v/b_3)^{1/2}$ for v>0. This is a kind of bifurcation, which is named critical.

However, when b₃ is negative, this solution does not exist anymore, and one shall include fifth order term,…, in the expansion. This generates a new possible solutions $u_2$ for u in the vicinity of v=0, **with a domain of existence defining $v_o$: $v_o < v < 0$**. The solution $u_1=0$ remains valid; and $u_2$ varies continuously with v. When v becomes positive, the solution $u_1=0$ is no more valid and the system jumps to $u_2$; the evolution of u presents a discontinuity at v=0; the solution jumps from $N_1=N_2=N<N_m$ to $N_1 \neq N_2$ when $N>N_m$. This is exemplified in Fig. A1.f. It is in general associated with hysteresis because if one reduces v below v=0, the solution $u_2$ evolves continuously till the system reaches $v_o$. When the system crosses $v_o$, $u_2$ does not exist anymore and the system jumps to u=0, which implies $N_1=N_2=N<N_m$. Such a bifurcation with a jump is called sub-critical. The general shape of j(N) which corresponds to this case is given in Fig. A1.e and the solutions are sketched in Fig. A1.f (pink curve); this curve fixes the value of j of the compartments, which fixes in turn the values $N_1$ and $N_2$ or $N=(N_1+N_2)/2$: When pink curve falls on the j curve of Fig. A1.e, it means that the two compartments have the same population $N_1=N_2$; when j does not fall on this curve of Fig. A1.e, it means that the two compartments have different populations $N_1 \neq N_2$; the two populations have to respect $j(N_1)=j(N_2)$ to be at equilibrium; so one population is larger than $N_m$, the other one smaller than $N_m$. The red dashed line in Fig. A1.f corresponds to unstable solutions with $N_1=N_2$.

So, it is worth noting that the situation is completely controlled by the shape of j(N) and by the evolution of the coefficient $b_3$ with the parameter of vibration.

A special situation is encountered when the curve j is fully symmetric compared to a vertical plane passing by point (j=0, $N=N_m$). It has been found to generate a special bifurcation called the **_hyper-critical bifurcation [11]_**. In this case, the solution $N_1=N_2=N$ is valid and stable for $N<N_m$. But no solution occurs for $N>N_m$, and a continuous set of solutions are found for $N=N_m$: all values of $N_1$, in the range $0<N_1<2N_m$ with $N_2=2N_m-N_1$ are possible. It means in this case that the grain number will evolve spontaneously in each box due to the noise existing on the flows $j_1$ and $j_2$; furthermore, as experimental conditions impose $N_1+N_2=2N$, it imposes $j_1-j_2=0$ in turn. This noise will generate the random walk of the population $N_1$ (and its symmetric $N_2$) as described in [11]. If the noise on $j_1$ depends on the number of grains in box 1, the jumping rate will depend on $N_1$ and the probability distribution at equilibrium will not be uniform but will depend on $N_1$ (see [11]).

This situation is not observed in the experimental case of the "Maxwell's daemon in granular gas" because it requires all $b_{2k+1}=0$, i.e. all $d^{2k+1}j/dN^{2k+1})_{at\ N=N_m}=0$, while the experimental data demonstrates only that $b_3=0$ when passing at 36 Hz in our experiment and $b_3=0$ when passing at 55Hz in the case of [10].

When $b_3=0$ it is expected that the noise plays more important influence. This is because the convergence towards the equilibrium state is looser there. A way to see this point is to analyse the dynamics using an expansion in the vicinity of the working point N, and not near $N_m$. This is the normal procedure; this is what is done now. Be $2N=N_1+N_2$, we label $\varepsilon=N_1-N=N-N_2$. So $\varepsilon=u/2$ of the preceding subsection. The dynamics obey then:

$$d\varepsilon/dt=j(N-\varepsilon)-j(N+\varepsilon) \tag{A7}$$

Expanding j(N) in Taylor series leads to the dynamics:

$$d\varepsilon/dt = -2 \ \{\varepsilon\ j^{(1)}(N)+ \varepsilon^3 j^{(3)}(N)/3!+ \varepsilon^5 j^{(5)}(N)/5!+....\} \tag{A8}$$

where $j^{(k)}(N)=d^kj/dN^k$ at point N; so Eq. (A8) depends on impairs terms in $\varepsilon$ only; it is different from Eq. (A4). Limiting to leading order, it results then the following from Eq. (A8):

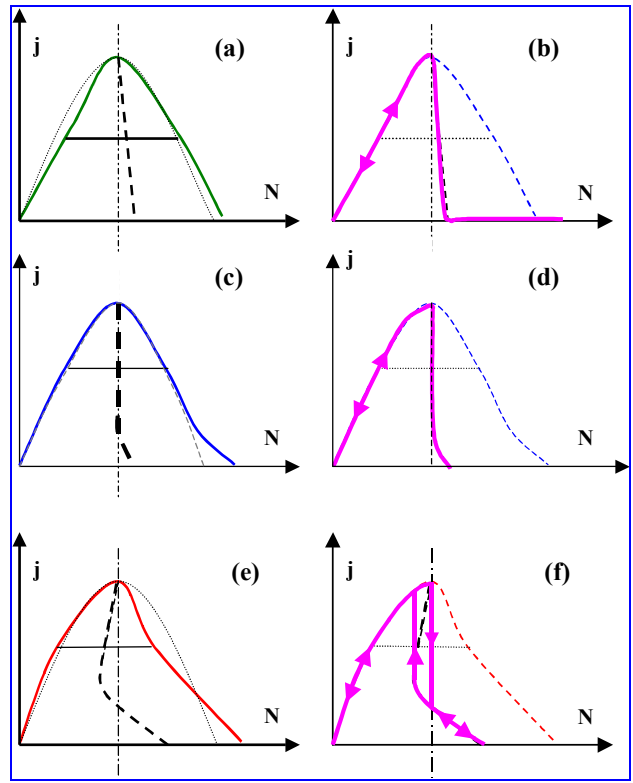

**Figure A1: The different possible configurations depending on the shape of the flow j(N).** *Curves made of points in Figs. a, b, c represents symmetric j(N). The dashed curves in these Figs correspond to locations of N such as N=(N₁+N₂)/2 with N₁≠N₂ and j(N₁)=j(N₂)*

*When j(N) is symmetric (Figs. c & d), it leads to the solutions (curve in pink in Fig. d);it consists in (i) a continuous set of possible j₁=j₂ when N=Nₘ , (ii) a unique possibility j₁=j₂ for N<Nₘ and (iii) no possibility when N>Nₘ.*

*In the case of Fig. a, j(N) is such that the slope of j vs. N is steeper before the maximum (at Nₘ) than after it. Then it exists a single equilibrium solution N₁=N₂=N for N<Nₘ; and a twofold solution with j₁=j₂ but N₁≠N₂ when N₁+N₂>Nₘ. When written as a function of the mean N=(N₁+N₂)/2, the flow curve j is singly valuated ; it is the curve in pink in Fig. b; below Nₘ. Equilibrium requires N₁=N₂=N when N<Nₘ, while it imposes N₁≠N₂ above Nₘ; this generates a bifurcation at N=Nₘ; as the difference N₁-N₂ quite (infinitely) fast with N-Nₘ just above Nₘ, the bifurcation is called critical.*

*In the case of Fig. e, the slope of j vs. N is steeper above the maximum (at N>Nₘ) than before it; the solutions are drawn in pink in Fig. f : it exists only a single solution satisfying j₁=j₂ when N<Nₒ<Nₘ, it imposes N₁=N₂ ; this solution remains stable till N<Nₘ . But a second set of solution, which is twofold, occurs in the vicinity of Nₘ , i.e. in the range Nₒ<N<Nₘ ; it is characterised by N₁≠N₂ but j₁=j₂ . Above (N₁+N₂)/2=N>Nₘ , a single twofold solution remains at this stage imposing N₁≠N₂. This shape of j(N) generates a bifurcation at N=Nₘ, of the sub-critical kind, i.e. with a jump, with hysteretic behaviour in between Nₒ and Nₘ.*

*In the experiment with vibro-fluidised granular gas, case a & e are observed. The fully symmetric case c is not demonstrated; also case c is not required to go continuously from case a to case e. As shown in the text passing from behaviour a to behaviour e requires only that coefficient b₃ of Eq. (A3.a) passes from positive (behaviour a) to negative (behaviour e) where as symmetry of behaviour c requires all (d²ᵏ⁺¹j/dN²ᵏ⁺¹)ₐₜ ₙ₌ₙₘ =0. Conversely b₃=0 imposes (d³j/dN³)ₐₜ ₙ₌ₙₘ =0 simply . So complete symmetry of Fig. c is a much too restrictive condition.*

If $j^{(1)}(N)>0$, *i.e.* j increases with N (this is the case N<Nₘ of Figs. A1), the solution ε=0 is stable, and relaxation is exponential. Introducing next order term, *i.e.* $j^{(3)}(N)$, is possible: when $j^{(3)}(N)$ is positive this does not lead to new solutions; if $j^{(3)}(N)$ is negative, there might be also two other solutions ε±=±{- 6 $j^{(1)}/j^{(3)}\}^{\frac{1}{2}}$ *but they are unstable*. To get new stable solutions one shall introduce $j^{(5)}(N)$, but we will not do it now; it will lead to case of Figs. 2.e & 2.f. And so on.

If $j^{(1)}(N)<0$, *i.e.* j decreases when N increases (this is the case when $N>N_m$ in the case of Figs. A1.a, A1.c and A1.e ), the solution $\varepsilon=0$ becomes unstable. One has to include next order term. This leads to the new solutions $\varepsilon_{\pm}=\pm\{-6\ j^{(1)}/j^{(3)}\}^{\frac{1}{2}}$, which exists if $j^{(3)}(N)$ is positive; these two solutions are now stable. This leads to the configuration exemplified in Fig. A1.b. However, if $j^{(3)}(N)$ is negative, these two new solutions do not exist and development to fifth order or further is required to get a stable solution.

So, when $j^{(1)}(N)<0$ and $j^{(3)}(N)<0$, the solution $\varepsilon=0$ is unstable and the fifth order term has to be included. When it is included, it leads to 2 new pairs of solutions if $\Delta^2=[j^{(3)}]^2/36-[\ j^{(1)}j^{(5)}]/30>0$, which are

$$\varepsilon_{1\pm} = [-10\ j^{(3)}/j^{(5)} \pm 60\Delta/j^{(5)}]^{\frac{1}{2}} \qquad (A9)$$

When $j^{(1)}(N)=0$, the convergence of the dynamics is governed by $j^{(3)}(N)$. It means first that relaxation towards $\varepsilon=0$ is no more exponential (relaxation time $\tau$ becomes infinite) but it follows a power law (critical slowing down). Furthermore when $j^{(3)}(N)=0$, the convergence is even looser, and the disturbance generated by some existing noise will become stronger.

These trends are summed up in Table 1.

| | v<0 | v>0 | bifurcation |
|---|---|---|---|
| $b_3>0$ , $b_4>0$ | u=0 | $u_{\pm}=\pm(4vb/c+5v^2+...)^{\frac{1}{2}}$ | critical |
| $b_3>0$ , $b_4<0$ | u=0 or | | Sub-critical |

**Table 1:** *Kind of evolution and bifurcation when* j(N) *is expanded in the vicinity of its maximum:* j(N)= $\Sigma_k\ b_k\ (N-N_m)^k$. ($b_1=0$, $b_2<0$ *because j is maximum at* $N_m$)

| | $N<N_m\ =>\ \ j^{(1)}>0$ | $N>N_m\ =>\ \ \ j^{(1)}<0$ | bifurcation |
|---|---|---|---|
| $j^{(3)}>0$ | $\varepsilon=0$ | $u_{1,2}=\pm(4vb/c+5v^2+...)^{\frac{1}{2}}$ | critical |
| $j^{(3)}<0$, $j^{(5)}>0$ | $\varepsilon=0$ or $\varepsilon\neq0$ | $\varepsilon\neq0$ | Sub-critical |

**Table 2:** *Evolution of the system as a function of the coefficient of the Taylor series of* j(N+$\varepsilon$)= $\Sigma_k$ $j^{(k)}/k!\ \varepsilon^k$, *around* N. *Bifurcation occurs when* $j^{(1)}$ *crosses from positive, to negative*.

### *A.3.  Examples for* **j(N)**

In [1], the flow was assumed to vary as $\qquad$ j= $j_o$ $N^2$ exp(-$\alpha N^2$) .
This leads to dj/dN= 2N(1-$\alpha N^2$) exp(-$\alpha N^2$), so that bifurcation occurs at $N_m=(1/\alpha)^{\frac{1}{2}}$ . This predicts also $d^3j/dN^3$= $j_o$ $\alpha N\{-20+32\alpha N^2-8(\alpha N^2)^2\}$ exp(-$\alpha N^2$), so that it predicts a critical bifurcation, *i.e.* $[d^3j/dN^3]_{N_m}$ = $j_o$ $4\alpha N_m$ exp(-1). However this equation predicts that j increases as $N^2$ near N=0, which is not observed experimentally [9].

$\qquad$ If one uses $\qquad\qquad\qquad\qquad$ j= $j_o$ N exp(-$\alpha N$)
which varies linearly with N near N=0 and which exhibits a maximum. The derivatives of this function are:

$$\frac{d^k j}{dN^k}= j^{(k)}(N)= (-\alpha)^{k-1}\ j_o\ \{k-\alpha N\}\ exp(-\alpha N) \qquad (A10)$$

$\qquad$ Consequently the maximum of j occurs at $N_m$= 1/$\alpha$ , $b_2$=-($d^2j/dN^2$)$_{N_m}$= $j_o$ $\alpha/e$ , $b_3$=2$\alpha^2$ $j_o/e$,…. As $b_3>0$ in this case, one shall observe a critical bifurcation at $N_m=1/\alpha$ . This does not allow the bifurcation to change of nature.

$\qquad$ Let us now assume that $\alpha$ varies with the acceleration $\Gamma$ and write

$$\alpha = \alpha_o + \beta\Gamma \tag{A11}$$

And let us assume that $\beta < 0$. Then this may describe the experimental results: indeed for a given set of N in the range $1/\alpha_o < N < 1/(\alpha_o + \beta\Gamma_{max})$ the system undergoes a critical bifurcation when lowering $\Gamma$. The tri-critical bifurcation occurs at

$$\Gamma_c = [1/N - \alpha_o]/\beta \tag{A12}$$

This might describe roughly what is occurring during the experiment.